\documentclass[twocolumn, tighten, times]{aastex631}

\usepackage{graphicx}
\usepackage{epstopdf}
\usepackage{float} 
\usepackage{bm}

\usepackage{color}
\usepackage{soul}

\DeclareUnicodeCharacter{02BC}{}

\shorttitle{COSMOS-DASH}
\shortauthors{Song.}
\graphicspath{{./}{figures/}}

\received{Dec 7, 2022}
\revised{Apr 24, 2023}
\accepted{May 1, 2023}

\begin{document}
\title{THE EFFECT OF ENVIRONMENT ON THE PROPERTIES OF THE MOST MASSIVE GALAXIES AT $0.5<z<2.5$ IN THE COSMOS-DASH FIELD}

\author[0000-0002-0846-7591]{Jie Song}
\affil{Deep Space Exploration Laboratory / Department of Astronomy, University of Science and Technology of China, Hefei 230026, China; \url{xkong@ustc.edu.cn}} 
\affil{School of Astronomy and Space Science, University of Science and Technology of China, Hefei 230026, People’s Republic of China}

\author[0000-0001-9694-2171]{GuanWen Fang}
\altaffiliation{GuanWen Fang and Jie Song contributed equally to this work}
\affil{Institute of Astronomy and Astrophysics, Anqing Normal University, Anqing 246133, Peopleʼs Republic of China; \url{wen@mail.ustc.edu.cn}} 

\author[0000-0003-3196-7938]{Yizhou Gu}
\affil{School of Physics and Astronomy, Shanghai Jiao Tong University, 800 Dongchuan Road, Minhang, Shanghai 200240, People’s Republic of China} 

\author[0000-0001-8078-3428]{Zesen Lin}
\affiliation{Department of Physics, The Chinese University of Hong Kong, Shatin, N.T., Hong Kong S.A.R., China}
\affil{Deep Space Exploration Laboratory / Department of Astronomy, University of Science and Technology of China, Hefei 230026, China; \url{xkong@ustc.edu.cn}} 
\affil{School of Astronomy and Space Science, University of Science and Technology of China, Hefei 230026, People’s Republic of China}

\author[0000-0002-7660-2273]{Xu Kong}
\affil{Deep Space Exploration Laboratory / Department of Astronomy, University of Science and Technology of China, Hefei 230026, China; \url{xkong@ustc.edu.cn}} 
\affil{School of Astronomy and Space Science, University of Science and Technology of China, Hefei 230026, People’s Republic of China}

\begin{abstract}
How the environment influences the most massive galaxies is still unclear. 
To explore the environmental effects on morphology and star formation in the most massive galaxies at high redshift, we select galaxies with stellar mass $\log(M_{\star}/M_{\odot})>11$ at $0.5<z<2.5$ in the COSMOS-DASH field, which is the largest field with near-infrared photometrical observations using HST/WFC3 to date. Combining with the newly published COSMOS2020 catalog, we estimate the localized galaxy overdensity using a density estimator within the Bayesian probability framework. With the overdensity map, no significant environmental dependence is found in the distributions of S\'{e}rsic index and effective radius. When we consider the star formation state, galaxies in lower density are found to have higher median specific star formation rate (sSFR) at $0.5<z<1.5$. But for star-forming galaxies only, sSFR is independent of the environment within the whole redshift range, indicating that the primary effect of the environment might be to control the quiescent fraction. Based on these observations, the possible environmental quenching process for these massive galaxies might be mergers.
\end{abstract}

\section{Introduction} \label{sec:intro}
Galaxies can be broadly classified into two types according to their star formation state and morphology: quiescent galaxies (QGs) with little to no ongoing star formation activity, relatively red colors, and spheroid-dominated morphologies; and star-forming galaxies (SFGs) with active star formation activity, fairly blue colors, and disk-dominated morphologies(e.g., \citealt{Strateva_2001,kauffmannStellarMassesStar2003,baldryQuantifyingBimodalColor2004,bellGalaxyBulgesTheir2008a,vandokkumFIRSTRESULTS3DHST2011,schawinskiGreenValleyRed2014,guMorphologicalEvolutionAGN2018}). In the past decade, it has been well-studied that these galaxy types are related to the local density (i.e., the galaxy environment) and stellar mass ($M_{\star}$) of galaxies. Quiescent, spheroidal galaxies are preferentially found in denser environments with a larger stellar mass, while SFGs are preferentially found in fields with a smaller stellar mass (e.g., \citealt{kauffmannEnvironmentalDependenceRelations2004a, blantonPhysicalPropertiesEnvironments2009,pengMASSENVIRONMENTDRIVERS2010b,
wooDependenceGalaxyQuenching2013}).

Physical mechanisms that quench galaxies can be broadly classified into two types. The one is often referred to as ``mass quenching'', which is mainly driven by internal physics \citep{pengMASSENVIRONMENTDRIVERS2010b}.
Possible mechanisms are feedbacks from active galactic nuclei and supernovae, which cease the star formation in galaxies by heating, expelling, and consuming gas (e.g., \citealt{birnboimBurstingQuenchingMassive2007,knobelQUENCHINGSTARFORMATION2015,terrazasQUIESCENCECORRELATESSTRONGLY2016}).
Besides internal physics, the external environment in which galaxies reside is expected as another crucial factor that can influence galaxy evolution. 
Higher-density environments can lead to enhanced merger rates, which may lead to a fast quenching process and change the shape and SFR of galaxies in a short timescale (e.g., \citealt{pengMASSENVIRONMENTDRIVERS2010b,faisstConstraintsQuenchingMassive2017}).

Observations show that internal physics and the external environment can affect galaxy formation and evolution. In the local universe, star formation activities of galaxies are closely related to their environments (e.g., \citealt{vanderwelEvolutionFieldCluster2007a}). The separable effects of stellar mass and environment on galaxy properties are discussed out to $z\sim 3$ (e.g., \citealt{pengMASSENVIRONMENTDRIVERS2010b,quadriTRACINGSTARFORMATIONDENSITYRELATION2012,muzzinGEMINICLUSTERASTROPHYSICS2012a,leeEVOLUTIONSTARFORMATION2015,darvishEFFECTSLOCALENVIRONMENT2016}). 
The denser environments generally tend to make galaxies older, redder, more spheroidal, and less star-forming. Galaxy environments are believed to have more significant impacts on low-mass galaxies ($M_{\star} < 10^{10}~M_{\odot}$; e.g., \citealt{leeEVOLUTIONSTARFORMATION2015,paulino-afonsoVISCOSSurvey2018,sobralDependenceStarFormation2010}). For the most massive galaxies ($M_{\star} \gtrsim 10^{11}~M_{\odot}$), the environments seems to have trivial effects on galaxy star formation state (e.g., \citealt{kawinwanichakijEffectLocalEnvironment2017,leeEVOLUTIONSTARFORMATION2015,paulino-afonsoVISCOSSurvey2018}). 
For instance, \cite{leeEVOLUTIONSTARFORMATION2015}, with a sample from UDS and SXDS, found that the difference in the quiescent fraction of galaxies with $M_{\star}>10^{10.5}~M_{\odot}$ between cluster and field environments is negligible at $0.5<z<2$.

However, recent studies reported that star formation of the most massive galaxies is also influenced by the environment (e.g., \citealt{Chartab_2020,darvishEFFECTSLOCALENVIRONMENT2016}). 
\cite{darvishEFFECTSLOCALENVIRONMENT2016} found that the most massive galaxies at $z<1$ in the COSMOS field have a smaller median star formation rate (SFR) in the denser environment. Still, at higher redshift, such median SFR seems independent of the environment. 
In \cite{Chartab_2020}, galaxy environments are found to have a significant effect on the median SFR for galaxies with $\log(M_{\star}/M_{\odot})>11$ in the CANDELS fields up to $z \sim 3$. From the observational point of view, the environmental effects on the properties of the most massive galaxies are not yet well understood. 

Another entry point to investigate the environmental effects in galaxy evolution is galaxy morphology. The environment of galaxies in the local universe is supposed to influence the most massive galaxy's morphologies (e.g., \citealt{dresslerGalaxyMorphologyRich1980,gotoMorphologydensityRelationSloan2003,skibbaGalaxyZooDisentangling2009}). If minor mergers dominate the size growth of galaxies, then galaxies might appear larger at fixed stellar mass in denser environments where mergers are more efficient, at least at lower redshifts.
\cite{yoonMASSIVEGALAXIESARE2017} used a sample from SDSS and showed that massive galaxies in higher density are found to have a larger size. Similar results have also been revealed for galaxies at higher redshift (e.g., \citealt{cooperDEEP3GalaxyRedshift2011,papovichCANDELSOBSERVATIONSSTRUCTURAL2012}).
Based on a sample from UDS, \cite{laniEvidenceCorrelationSizes2013} found that the most massive galaxies ($M_{\star}>2\times10^{11} M_{\odot}$) at $1<z<2$ in the higher density environments are typical 50\% larger in size compared to those in the lowest density environments. However, some observations lead to a conflicting result that environments do not affect galaxy size (e.g., \citealt{weinmannEnvironmentalEffectsSatellite2009,maltbyEnvironmentalDependenceStellarmasssize2010,retturaFORMATIONEPOCHSSTAR2010,huertas-companyNOEVIDENCEDEPENDENCE2013}), or even a negative effect (e.g., \citealt{raichoorEARLYTYPEGALAXIESIV2012}).

To further study whether there is an environmental dependence of galaxy properties for the most massive galaxies and when the dependence starts (if any), we construct a large sample and the overdensity map at $0.5<z<2.5$ using the COSMOS2020 catalogs \citep{weaverCOSMOS2020PanchromaticView2022}. To extend the morphological analysis at higher redshift, we apply both parametric and nonparametric measurements on galaxy images from the COSMOS-DASH field, which is the largest area near-infrared (NIR) survey with the Wide Field Camera 3 (WFC3) on board HST to data. Utilizing this sample, we explore how the local environment of the most massive galaxies affects galaxy morphologies, sSFR, and quiescent fraction out to $z\sim2.5$.

The layout of this paper is as follows. In Section \ref{sec:data}, we review the basic data from the COSMOS-DASH program and give the criteria we use to select the most massive galaxies. In Section \ref{sec:morph para}, we introduce the morphology parameters used in this paper. We present the mass and environment distributions in Section \ref{sec:distribution}. The results of the environmental effect on morphologies and star formation activities are shown in Sections \ref{sec: morph effect} and \ref{sec:strformation effect}, respectively. The conclusions are presented in Section \ref{sec:summary}. Throughout the paper, we adopt a flat $\Lambda$ cold dark matter ($\Lambda$CDM) cosmology with $H_0 = 70~\mathrm{km~s^{-1}~Mpc^{-1}}$, $\Omega_{\rm{m}}=0.3$, and $\Omega_{\rm{\Lambda}}=0.7$, and a \cite{chabrierGalacticStellarSubstellar2003} initial mass function.

\section{Data and Sample selection} \label{sec:data}

\subsection{The COSMOS2020 Catalog} \label{subsec:catalog}

Our galaxy sample is constructed from the COSMOS2020 catalog of \cite{weaverCOSMOS2020PanchromaticView2022}, a new reference photometric catalog across the $2\ \rm deg^2$ COSMOS field. The catalog is detected from the ``chi-squared'' izYJHKs image, which is created with SWarp \citep{Bertin2002swarp} from the combined original images without PSF homogenization. The extensive and continuous coverage of the field would have a great benefit to the construction of an overdensity map. With a more reliable photometric estimation in a wide wavelength range from far-ultraviolet (FUV) to NIR, they fit the observed spectral energy distribution (SED) with galaxy models to estimate photometric redshift and other galaxy physical properties. From this catalog, we extract the redshift, stellar mass, and other physical properties of our galaxy sample.

Two versions of photometric catalogs derived from different source extraction methods are provided by \cite{weaverCOSMOS2020PanchromaticView2022}. Photometric redshift is also estimated via two other codes EAZY \citep{Brammer_2008} and LePhare \citep{ilbertAccuratePhotometricRedshifts2006}. Figure 15 of \cite{weaverCOSMOS2020PanchromaticView2022} shows that photometric redshift derived from LePhare using the Farmer version of the photometric catalog (based on SEP and Tractor) is more reliable for faint sources. Due to the lack of spectroscopic redshift in this deep field, photometric redshift is very important in estimating the overdensity map. We thus adopt the LePhare redshift derived based on the Farmer version catalog in this work.

The fitting code Lephare was used with two different configurations to derive: (1) photometric redshift ($z_{\rm{LePh}}$); (2) stellar masses ($M_{\rm{LePh}}$) and SFR. In the first run of LePahare, a library of 33 galaxy templates extracted from \cite{ilbertCOSMOSPHOTOMETRICREDSHIFTS2009} and \citet[hereafter BC03]{bruzualStellarPopulationSynthesis2003} is used to fit the observed SED.
Several options of dust extinction/attenuation curves (e.g., the starburst attenuation curve of \citealt{calzettiDustContentOpacity2000}, the SMC extinction curve of \cite{1984A&A...132..389P}, and two variations of Calzetti law including the 2175 \AA\ bump) are used to modify the intrinsic spectrum of these templates. The photometric redshift $z_{\rm{LePh}}$ is defined as the median of the redshift likelihood function after evaluating the $\chi^2$ of each fit. After fixing the redshift of each source to $z_{\rm{LePh}}$, templates derived from the BC03 models were used to determine galaxy stellar mass in the second run. In this step, exponentially declining star formation histories (SFH) and delayed $\tau$ SFH with $\tau$ of 0.1 - 30 Gyr were considered.

\subsection{Star Formation Rates} \label{subsec:SFR}

SFR used in this work is estimated from the combination of the rest-frame UV luminosity ($L_{2800}$) and the total infrared luminosity ($L_{\rm IR}$). With an assumption of $f_\lambda \propto \lambda^{\beta}$, the rest-frame 2800 \AA\ flux is computed by extrapolating the rest-frame GALEX FUV and NUV fluxes, which are the output of Lephare and provided by the COSMOS2020 catalog. As for the total infrared luminosity $L_{\rm IR}$, monochromatic extrapolation based on the \cite{Wuyts_2008} IR SED template using the observed Spitzer/MIPS 24 $\mu$m flux is proved to be a good estimator for galaxies within the redshift range explored in this work \citep{wuytsSTARFORMATIONRATES2011,linHOWACCURATEARE2016}. We thus adopt this extrapolation method to derive $L_{\rm IR}$ for our sample. Following \cite{whitakerCONSTRAININGLOWMASSSLOPE2014}, we then convert $L_{2800}$ and $L_{\rm IR}$ into the total SFR by
\begin{equation}
    \mathrm{SFR_{ tot}}[M_{\odot}~\mathrm{yr}^{-1}] = 1.09\times 10^{-10} (L_{\rm IR} + 3.3L_{2800})/L_{\odot}.
\end{equation}

Since the MIPS 24 $\mu$m fluxes are available for the vast majority of our sample (1669/1684), we argue that the exclusion of $<1\%$ galaxies that do not have 24 $\mu$m fluxes (and thus SFR estimation) should not have any significant effect on the statistical result presented in the following analysis. When mentioning SFR in the following part, we only consider those sources with MIPS 24 $\mu$m measurements.

\subsection{Images of COSMOS-DASH} \label{subsec:COSMOS-DASH}

Wide-field NIR surveys have been shown invaluable for studying the morphology of high-redshift galaxies since the rest-frame optical emission of these galaxies shifts into NIR. Some surveys have been undertaken from the ground (e.g., NMBS, \citealt{whitakerNEWFIRMMEDIUMBANDSURVEY2011}; UltraVISTA, \citealt{mccrackenUltraVISTANewUltradeep2012a}).  
Compared with the ground-based survey, the standard HST observations are able to obtain deeper images with higher resolution.  
In the standard HST observations, guide stars are required for each pointing, which limits the NIR surveys to a small field of view. As a result, the largest area in the $J_{125}$ and $H_{160}$ bands for CANDELS deep fields is $\sim 0.25 \rm deg^2$ \citep{groginCANDELSCOSMICASSEMBLY2011,koekemoerCANDELSCOSMICASSEMBLY2011}. Drift and Shift (DASH) is an efficient technique to balance resolution, depth, and area for the HST observation. It circumvents the problem by acquiring guide stars for only the first pointing and guiding with the three HST gyros for the rest pointings. With this DASH technique, \cite{mowlaCOSMOSDASHEvolutionGalaxy2019} presented a COSMOS-Drift And SHift (COSMOS-DASH) survey in the COSMOS field. It is taken with 57 DASH orbits in the F160W filter of WFC3 and covers an area of $0.49\ \rm deg^2$ ($0.7\ \rm deg^2$ when combined with archival data), which is much larger than that of the same field observed by CANDELS. Since the exposure is about 300 seconds per pointing, the $5\sigma$ source depth of the image is $H_{160}=25.1$ AB mag. 

To enlarge the sample size and extend the morphological analysis at higher redshift, we employ the final mosaic images of the COSMOS-DASH field, which is the largest area with NIR imaging using HST/WFC3 till now.\footnote{It is available from the COSMOS-DASH website of \url{https://archive.stsci.edu/hlsp/cosmos-dash/}}
The details of parametric and nonparametric measurements of galaxy morphology are described in Section \ref{sec:morph para}. 
Although the depth is slightly shallower than that of CANDELS, it is sufficient for the morphological measurements of the most massive galaxies (e.g., \citealt{momchevaNewMethodWidefield2017}; \citealt{mowlaCOSMOSDASHEvolutionGalaxy2019}).

\subsection{Selection of Galaxies for Analysis} \label{subsec:sample selection}

We start the sample selection from the newly published COSMOS2020 catalog \citep{weaverCOSMOS2020PanchromaticView2022}. At first, galaxies that might suffer contamination from nearby bright stars are removed by $\mathrm{lp_{type}}=1$ and $\mathrm{FLAG_{COMBINED}}=0$, which means the objects: (1) are a galaxy rather than a star; (2) are not near a bright star.

\begin{figure}[htb!]
\centering
\includegraphics[width=0.48\textwidth, height=0.48\textwidth]{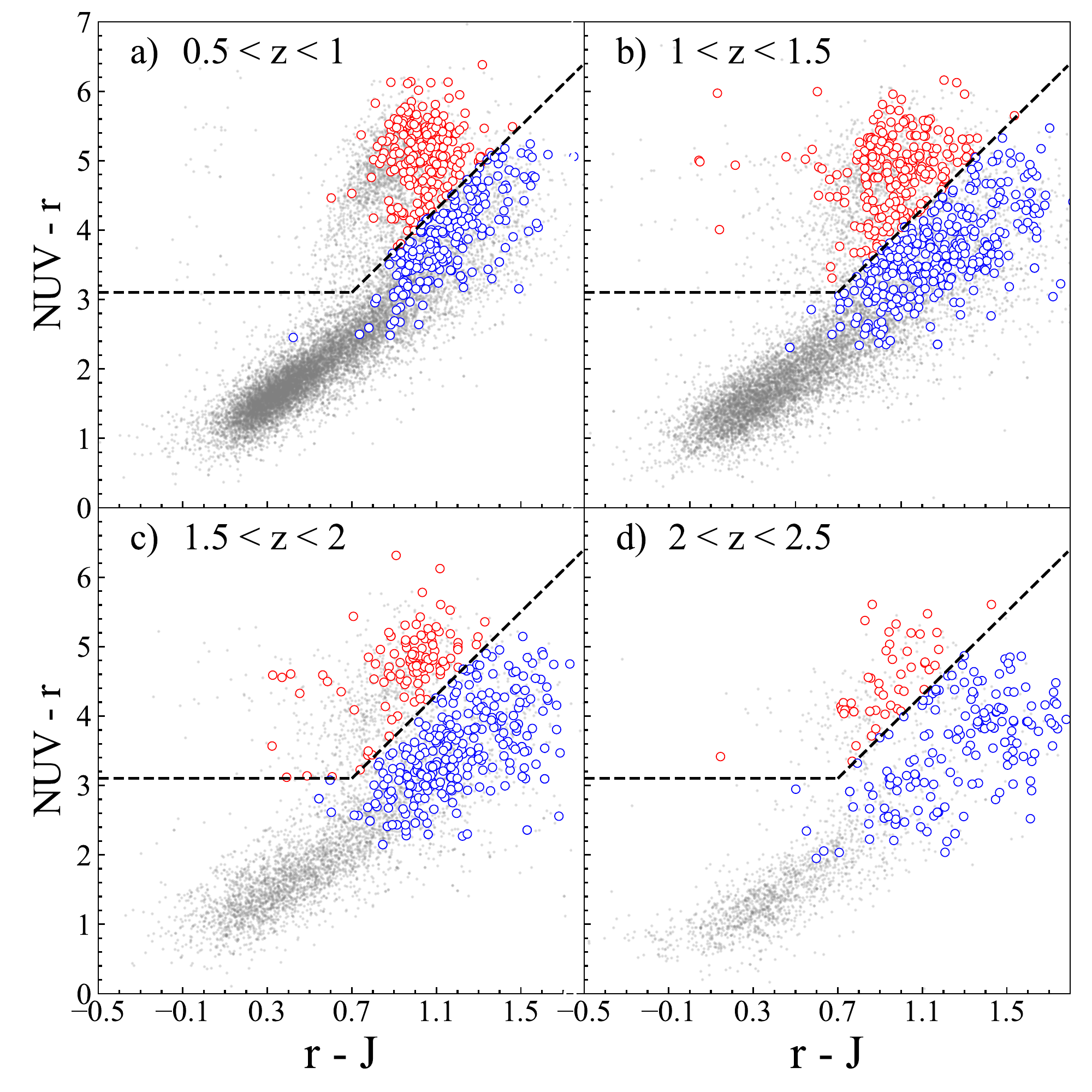}
\caption{Rest-frame $NUV-r$ vs. $r-J$ color distribution in four redshift bins. QGs and SFGs are separated by the selection criteria defined in Equations (\ref{eq1}) and (\ref{eq2}), shown by the black dotted lines. All the galaxies in the COSMOS-DASH field with $\log(M_{\star}/M_{\odot})>10$ in each redshift bin are denoted in grey. SFGs and QGs in our sample with $\log(M_{\star}/M_{\odot})>11$ are indicated by blue and red open circles, respectively.}
\label{fig1}
\end{figure}

To select the most massive galaxies, we limit our sample to galaxies with $\log(M_{\star}/M_{\odot})>11$ at $0.5<z<2.5$. QGs are separated from the star-forming population via the rest-frame $NUV-r$ and $r-J$ color selection \citep{weaverCOSMOS2020PanchromaticView2022}:
\begin{equation}
    NUV-r>3.1
\label{eq1}
\end{equation}
\begin{equation} 
    NUV-r>3(r-J)+1.
\label{eq2}
\end{equation} 
The distribution of our sample in the $NUV-r$ versus $r-J$ plane is shown in Figure \ref{fig1} in four redshift bins from z = 0.5 to 2.5. 
The blue and red open circles represent SFGs and QGs with $\log(M_{\star}/M_{\odot})>11$, respectively, while the light grey dots are all galaxies with $\log(M_{\star}/M_{\odot})>10$. From this figure, it is evident that the assembly of the quiescent population happens at late cosmic times since QGs are rare at $z > 2$ (e.g., \citealt{muzzinEVOLUTIONSTELLARMASS2013,tomczakGALAXYSTELLARMASS2014}). Given the above criteria, we finally obtain a sample of the most massive galaxies at $0.5<z<2.5$ that contains 1684 galaxies (including 1020 SFGs and 664 QGs) with the $H_{160}$ observations of COSMOS-DASH survey. 

\begin{figure}[htb!]
\centering
\includegraphics[width=0.48\textwidth, height=0.48\textwidth]{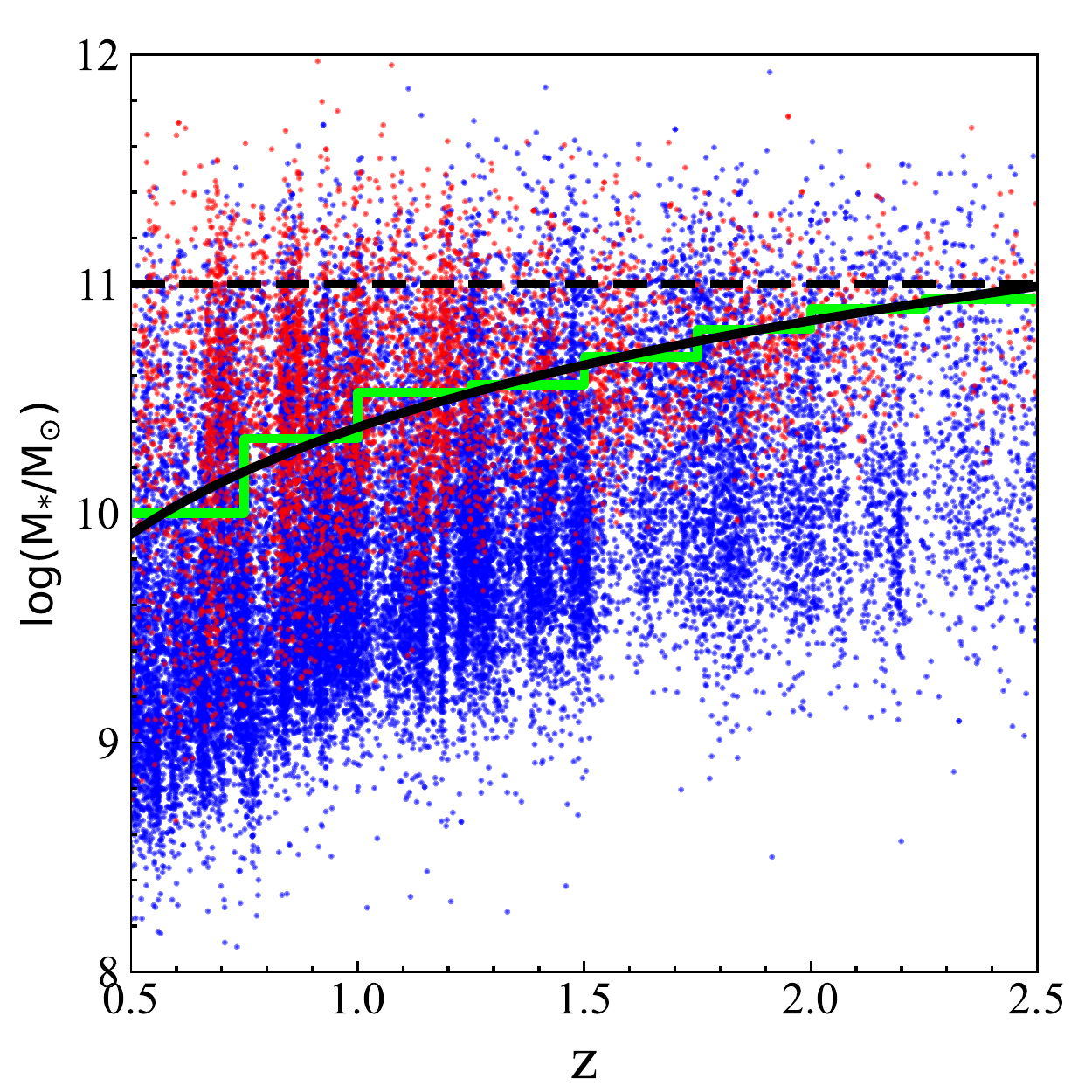}
\caption{Stellar mass as a function of redshift for the parent sample of this work. The red and blue dots represent star-forming and quiescent galaxies, respectively. The lime step-like line shows the mass completeness limits for QGs in the redshift interval $\Delta z=0.25$. The mass completeness limit can be parameterized with a function of $M_{\rm{comp}}(z)=10.375+0.671\ln(z)$ in this study, which has been shown by the black solid curve. The black dashed line is $M_{\star} = 10^{11}M_{\odot}$ that marks the mass lower limit of our sample.}
\label{fig2}
\end{figure}

In the magnitude-limited sample, stellar mass completeness depends on both redshift and mass-to-light ratio ($M/L$). QGs have a higher mass limit at a given magnitude limit than SFGs due to the larger $M/L$ ratios. 
The H-band limit $H_{\rm{lim}}$ in this work is 23.5 mag. 
To be conservative, we derive the mass limit only using the quiescent population. The mass completeness of QGs is estimated using the method described in \cite{pozzettiZCOSMOS10kbrightSpectroscopic2010}. 
The faintest 20\% of QGs are considered to derive a representative limit for our sample, and the completeness limit ($M_{\rm{comp}}$) is estimated within an interval of $\Delta z$ = 0.25. With the typical mass-to-light ratio for each galaxy, the stellar mass limit $M_{\rm{lim}}$ at a redshift slice is the mass that galaxies would have if their apparent magnitudes were equal to the magnitude limit. Specifically, the stellar mass limit at a specific redshift can be derived by $\log(M_{\rm{lim}})=\log(M_{\star}) +0.4(H-H_{\rm{lim}})$. 
Then $M_{\rm{comp}}$ is defined as the upper envelope of the $M_{\rm{lim}}$ distribution below which 90\% of the $M_{\rm{lim}}$ values locate at a given redshift. The mass completeness limits of these redshift bins can be parameterized as a function of redshift, $M_{\rm{comp}}(z)=10.375+0.671\ln(z)$, which describes how the mass completeness limits vary from $z = 0.5$ to 2.5 in this work.

In Figure \ref{fig2}, we exhibit the $M_{\star}$ distribution as a function of redshift for galaxies at $0.5 < z < 2$ in the COSMOS2020 catalog (i.e., the parent sample of this work). The measured and parameterized functions of $M_{\rm{comp}}(z)$ are denoted as the lime step-like line and black solid curve, respectively. From Figure \ref{fig2}, it is clear that our sample with stellar mass $\log(M_{\star}/M_{\odot})>11$ are all above the mass completeness limit.

Due to mergers and star formation, there may be significant differences in number density between high-redshift and low-redshift for these massive galaxies. When considering only a constant stellar mass threshold to select samples, the results may be affected by the ``progenitor bias'' and the ``inconsistency in terms of abundance'' at different redshift when interpreting the results across cosmic time. However, we are concerned with the difference between the nature of galaxies in a high-density environment and a low-density environment in each redshift bin, so we only need to care about the ``progenitor bias'' and the ``inconsistency in terms of abundance'' in each redshift bin. One way to mitigate this issue is to assume that galaxies maintain rank order across cosmic time. This assumption is generally consistent with the so-called ``down-sizing'' phenomenon and predicts a evolving co-moving number-density with redshift to trace the mass and size growth of galaxies when mergers are considered(e.g., \citealt{hillMassColorStructural2017}). Similar to \cite{hillMassColorStructural2017} and \cite{behrooziUSINGCUMULATIVENUMBER2013}, we calculate the difference in stellar mass by assuming an evolving co-moving density at a redshift interval of 0.5, which corresponds to the bin size of the redshift we choose in the following study. The result shows that the differences are all less than 0.2 dex. Considering that the mean error of the stellar mass obtained by SED fitting is about 0.25 dex in the COSMOS2020 catalog, we argue that such a small difference has no effect on the results in each redshift bin.

\section{Morphological Parameters} \label{sec:morph para}

\subsection{Band-Shifting}
The observed H-band samples the rest-frame V band at $z\sim2$, but at $z\sim 1$ it samples the rest-frame I band. Over this wavelength range, galaxies' morphological parameters might significantly change(e.g., \citealt{kennedyGalaxyMassAssembly2015, psychogyiosMultiwavelengthStructureAnalysis2020}). Due to this consideration, 
the COSMOS-DASH data is used only in the redshift range of $1.5< z<2.5$. For galaxies at $0.5<z<1.5$, we adopt the HST/ACS $I_{\rm 814}$ observations in the COSMOS field \citep{koekemoerCOSMOSSurveyHubble2007}.

\subsection{Parametric Measurements}

To describe the structure of galaxies at $1.5<z<2.5$ in our sample, we measured the S\'{e}rsic index $n$ and effective radius $r_{\rm e}$ by fitting the galaxy with a single S\'{e}rsic model with the $H_{\rm{160}}$ COSMOS-DASH image. GALFIT package \citep{pengDetailedStructuralDecomposition2002} is widely used in many studies to fit the surface brightness profile of galaxies with a form $\Sigma(r)=\Sigma_e {\rm exp} [-k(r/r_e)^{1/n}-1]$. 
In this study, we obtain the $r_e$ and n values for all galaxies at $1.5<z<2.5$ in the COSMOS-DASH field using the GALAPAGOS software \citep{bardenGalapagosPixelsParameters2012}, which is a wrapper of SExtractor and GALFIT for morphological analyses, with a position-dependent PSF applied. For galaxies at $0.5<z<1.5$, S\'{e}rsic index $n$ and effective radius $r_{\rm e}$ are extracted from the Advanced Camera for Survey Catalog \citep{griffithADVANCEDCAMERASURVEYS2012}, which includes approximately 470,000 astronomical sources derived from the AEGIS, COSMOS, GEMS, and GOODS fields. The morphological parameter is also obtained with the GALAPAGOS software by assuming a single s\'{e}rsic model.

\subsection{Nonparametric Measurements}

The nonparametric measurements of galaxy morphology do not assume a particular analytic function for the galaxy's light distribution and therefore are suitable for high redshift galaxies. We also calculate the nonparametric morphological parameters for all galaxies in our sample, based on the COSMOS-DASH $H_{\rm 160}$ observations at $1.5<z<2.5$ and the HST/ACS $I_{\rm 814}$ observations at $0.5<z<1.5$, to study how the environment affects galaxy morphology. In our analysis, we focus on the following indices.

\textbf{Gini coefficient}. It is originally used in economics to describe the wealth inequality of a population. \cite{abrahamNewApproachGalaxy2003} used it to indicate the distribution of a galaxy's flux. It can be computed as:
\begin{equation}
    G=\frac{1}{\overline{f}N_{\mathrm{pix}}(N_{\mathrm{pix}}-1)}\sum_{i=0}^{N_{\mathrm{pix}}}(2i-N_{\mathrm{pix}}-1)f_i
\end{equation}
Where $N_{\mathrm{pix}}$ is the number of pixels of the galaxy, $f_i$ is the pixel flux value sorted in ascending order, and $\overline{f}$ represents the mean over the pixel values.

\bm{$M_{20}$}. It is the normalized second-order moment of the brightest 20\% of the galaxy's flux. To compute $M_{20}$, the galaxy pixels are ranked in descending order by flux, and $M_{20}$ is defined as:
\begin{equation}
    M_{\rm{tot}}=\sum_{i}^{N_{\mathrm{pix}}} M_i=\sum_i^{N_{\mathrm{pix}}} f_i[(x_i-x_c)^2+(y_i-y_c)^2]
\end{equation}
\begin{equation}
    M_{\rm{20}}=\log_{10}\frac{\sum_i M_i}{M_{\mathrm{tot}}},\ \mathrm{while}\ \sum_i f_i=0.2f_{\mathrm{tot}},
\end{equation}
where $f_{\rm{tot}}$ is the total flux of the galaxy, $f_i$ is the flux value of each pixel i, $(x_i,\ y_i)$ is the position of pixel i, and $(x_c,\ y_c)$ is the center of the image. \cite{lotzNewNonparametricApproach2004} used $M_{20}$ to trace the spatial distribution of bright nuclei, bars, and off-center clusters. The Gini and $M_{20}$ are often used together to separate galaxies into different morphological subclasses (e.g., \citealt{Lotz_2006,rodriguez-gomezOpticalMorphologiesGalaxies2019}.

\cite{lotzNewNonparametricApproach2004} have shown that the reliability of nonparametric measurements is related to the S/N and the resolution of the image. Their work reported that Gini and $\rm M_{20}$ are reliable to within $\sim10\%$ for galaxy images with $\rm S/N >2$. Galaxies in our sample are bright enough that most of them have $\rm S/N > 5.0$, so the image depth should not have any significant effect on the measurements of the structural parameters.
Image PSF also has impact on the result. To simulate the effect of PSF, we select some galaxies and convolve the raw images with Gaussian kernels with different widths. The Gini coefficient and $\rm M_{20}$ index systematically decrease steadily as the image becomes less resolved. But it generally shows the offset is less than $10\%$ when the PSF of the image becomes twice as large. So we believe the nonparametric measurements in this work are reliable at whole redshifts.

\section{Mass and Environment Distribution} \label{sec:distribution}

\subsection{Mass Distribution}

Many previous studies showed that both stellar mass and the local environment have effects on galaxy properties(e.g., \citealt{darvishEFFECTSLOCALENVIRONMENT2016,kawinwanichakijEffectLocalEnvironment2017}). To study the environmental influence on galaxy properties at different redshift ranges, we first check whether there is a similar stellar mass distribution in different redshift bins to remove the possible impact of different stellar mass distributions.

\begin{figure}[htb!]
\centering
\includegraphics[width=0.48\textwidth, height=0.48\textwidth]{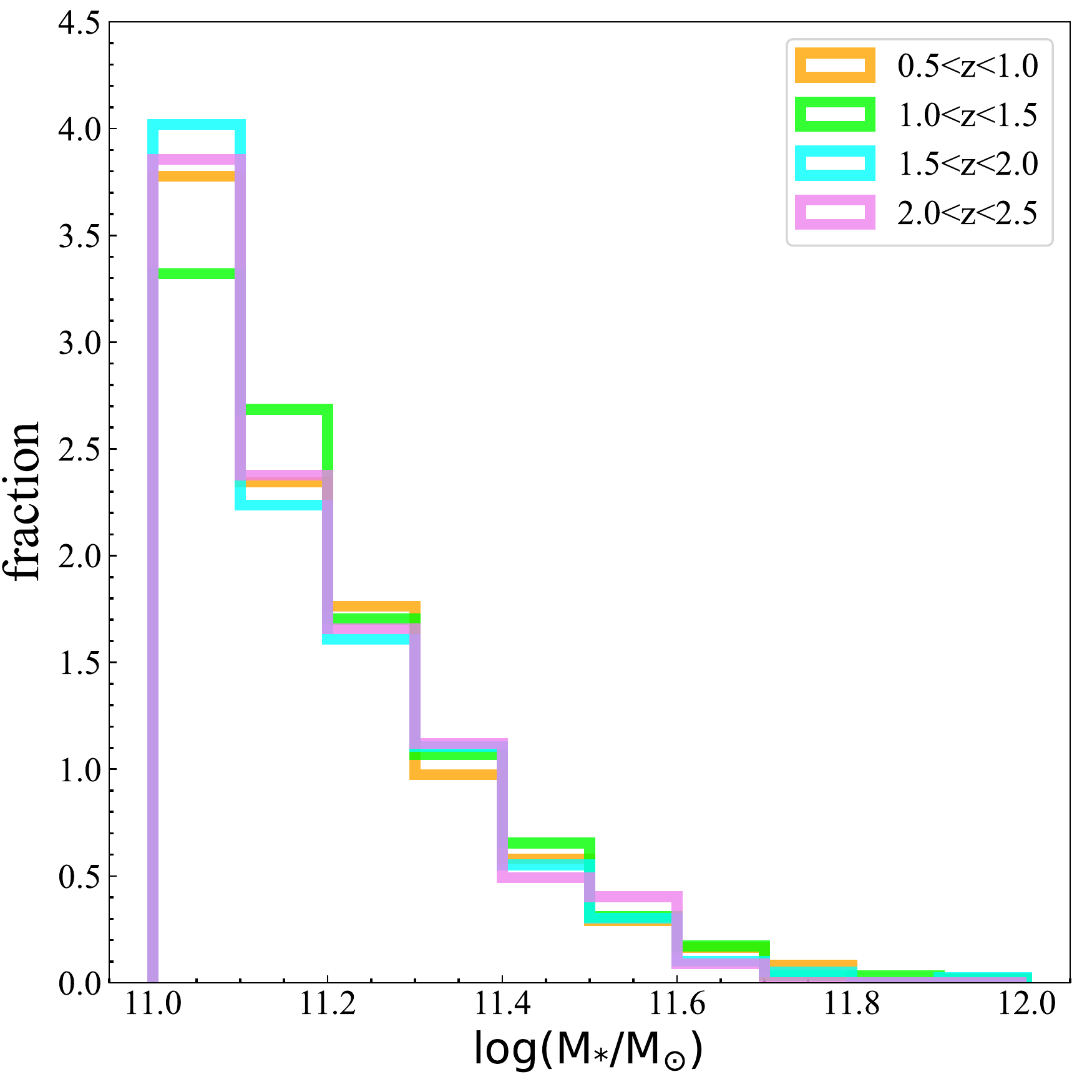}
\caption{Distribution of stellar mass in different redshift bins. Orange, lime, aqua, and violet histograms represent galaxy stellar mass distribution at $0.5<z<1.0$, $1.0<z<1.5$, $1.5<z<2.0$, and $2.0<z<2.5$ respectively.}
\label{fig3}
\end{figure}

The mass distributions of four redshift bins are shown in Figure \ref{fig3}. The median stellar masses in four redshift bins are $10^{11.14}~M_{\odot}$, $10^{11.15}~M_{\odot}$, $10^{11.13}~M_{\odot}$, and $10^{11.14}~M_{\odot}$ from the lowest redshift bin to the highest redshift bin, respectively. Obviously, galaxies in different redshift bins share a similar $M_{\star}$ distribution. Furthermore, we perform the Kolmogorov-Smirnov (K-S) test to quantify the statistical significance of the difference in the $M_{\star}$ distributions. The resulting $p$-values between the $M_{\star}$ distributions for galaxies at $0.5<z<1.0$ and those in the other three redshift bins are [0.177, 0.888, 0.897]. For galaxies at $1.0<z<1.5$, $1.5<z<2.0$, and $2.0<z<2.5$, the $p$-values of similar comparisons are [0.177, 0.051, 0.396], [0.888, 0.051, 0.788], and [0.897, 0.396, 0.788], respectively. The quality $p$-value describes the probability that two samples are drawn from the same underlying distribution. We adopt $p = 0.05$ as the upper limit probability to verify that the two subsamples have different $M_{\star}$ distributions. All the $p$-values listed above are more significant than the adopted 0.05 threshold, reaching a similar conclusion to the one revealed by Figure \ref{fig3}. According to this criterion, we believe galaxies in four redshift bins share the same stellar mass distribution.

\subsection{Environment Distribution}

The environment is a crucial external factor for galaxy evolution (e.g., \citealt{darvishCOMPARATIVESTUDYDENSITY2015,donnariQuenchedFractionsIllustrisTNG2020}). The density estimation method used in this paper has been described in detail in \cite{guEffectEnvironmentStar2021}. We briefly explain it in the following. The local density is estimated by $\Sigma_N' \propto 1/\Sigma_{i=1}^Nd_i^2$, where $d_i$ is the projected distance to the $i$th neighbour within a redshift slice ($|\Delta z| < 2\sigma_z(1+z),\sigma_z=0.02)$. In this work, the calculation of local overdensity involves all galaxies that are brighter than 26 mag in the IRAC1 band in the COSMOS2020 catalog within the individual redshift slice of $|\Delta z| < 2\sigma_z(1+z)$, including the galaxies with stellar mass less than $10^{11}M_{\odot}$.
Based on this local density map, we employ a dimensionless overdensity to measure the relative density of the environment, which is described as:
\begin{equation}
	1+\delta'_N = \frac{\Sigma_N'}{\langle\Sigma_N'\rangle_{\rm{uniform}}}=\frac{\Sigma_N'}{k'_N\Sigma_{\rm{surface}}}
\end{equation}
where $\langle\Sigma_N'\rangle_{\rm{uniform}}$ is the Bayesian density in uniform condition. The $\Sigma_{\rm{surface}}$ is the surface number density within a given redshift slice, and the $k_N'$ is the correction factor between $\langle\Sigma_N'\rangle_{\rm{}{uniform}}$ and $\Sigma_{\rm{surface}}$. In this paper, we adopt the local density defined by the five closest galaxies ($1+\delta_5'$) to calculate each target galaxy's local environmental density.

Some things to note are that the use of photometric redshifts to calculate environmental density may come with risks. Very large photo-z uncertainties tend to wash out the structures in the distribution of galaxies(e.g., \citealt{cooperMeasuringGalaxyEnvironments2005, malavasiReconstructingGalaxyDensity2016}). 
But in this study, due to a wealth of imaging and spectroscopic data that have been collected in the COSMOS field, the redshift accuracy in the COSMOS2020 catalog can reach a precision of 2\% for galaxies brighter than 26 mag in the IRAC1 band. Some studies such as \cite{laiCANWEDETECT2016} have shown that even larger uncertainties with $\sigma_z\leq0.06$ can still reveal the general environmentally driven trends. A similar test has also been taken by \cite{guEffectEnvironmentStar2021}, which has proven that our determination of overdensity, based on the Bayesian density estimator, is competent to trace the structures at high redshifts with $\sigma_{\Delta z/(1+z)}\sim0.02$. So, we think our overdensity map can reveal the general environmentally driven trends.

\begin{figure}[htb!]
\centering
\includegraphics[width=0.48\textwidth, height=0.48\textwidth]{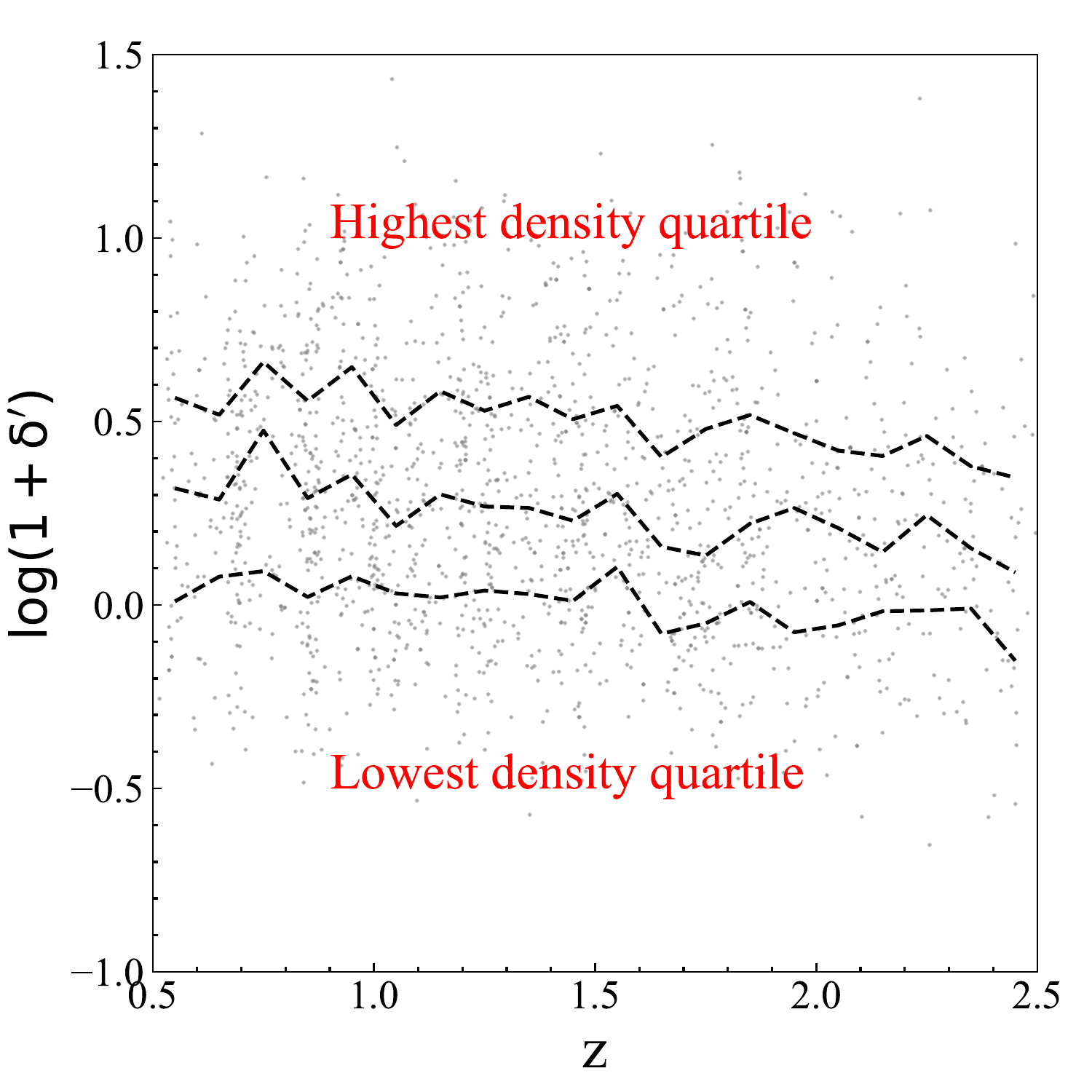}
\caption{The overdensity ($1+\delta'$) distribution as a function of redshift. The grey dots represent individual galaxies in our sample, while the black dashed curves indicate the 25\%, 50\%, and 75\% percentiles of the binned distributions, respectively, within a small redshift interval ($\Delta z=0.1$).}
\label{fig4}
\end{figure}

The distribution of the overdensities ($1+\delta_5'$) of the most massive galaxies is shown in Figure \ref{fig4}. To reduce the impact of galaxies in the intermediate density from which an ambiguous conclusion might be obtained, we divide the overdensity into four quarters as a function of redshift. The boundaries between adjacent quarters (i.e., the 25\%, 50\%, and 75\% percentiles of the binned distributions) are denoted as black dashed curves. From Figure \ref{fig4}, the median overdensity increases steadily with the cosmic epoch. This can be understood because almost all massive galaxies would migrate to higher overdensities as large-scale structure develops through gravitational instability. A similar result has also been unveiled by \cite{pengMASSENVIRONMENTDRIVERS2010b} at $0.1<z<1.4$.
\section{Environmental effect on morphology} \label{sec: morph effect}

In the $\Lambda$CDM models, the most massive galaxies are expected to form and grow through several major and minor mergers.  
The size of the most massive galaxies should depend on local density since more mergers are expected in higher density (e.g., \citealt{yoonMASSIVEGALAXIESARE2017,naabFormationEarlyType2007}). 
In this section, we will explore whether the environment affects galaxy morphology as a function of redshift.

\begin{figure}[htb!]
\centering
\includegraphics[width=0.5\textwidth]{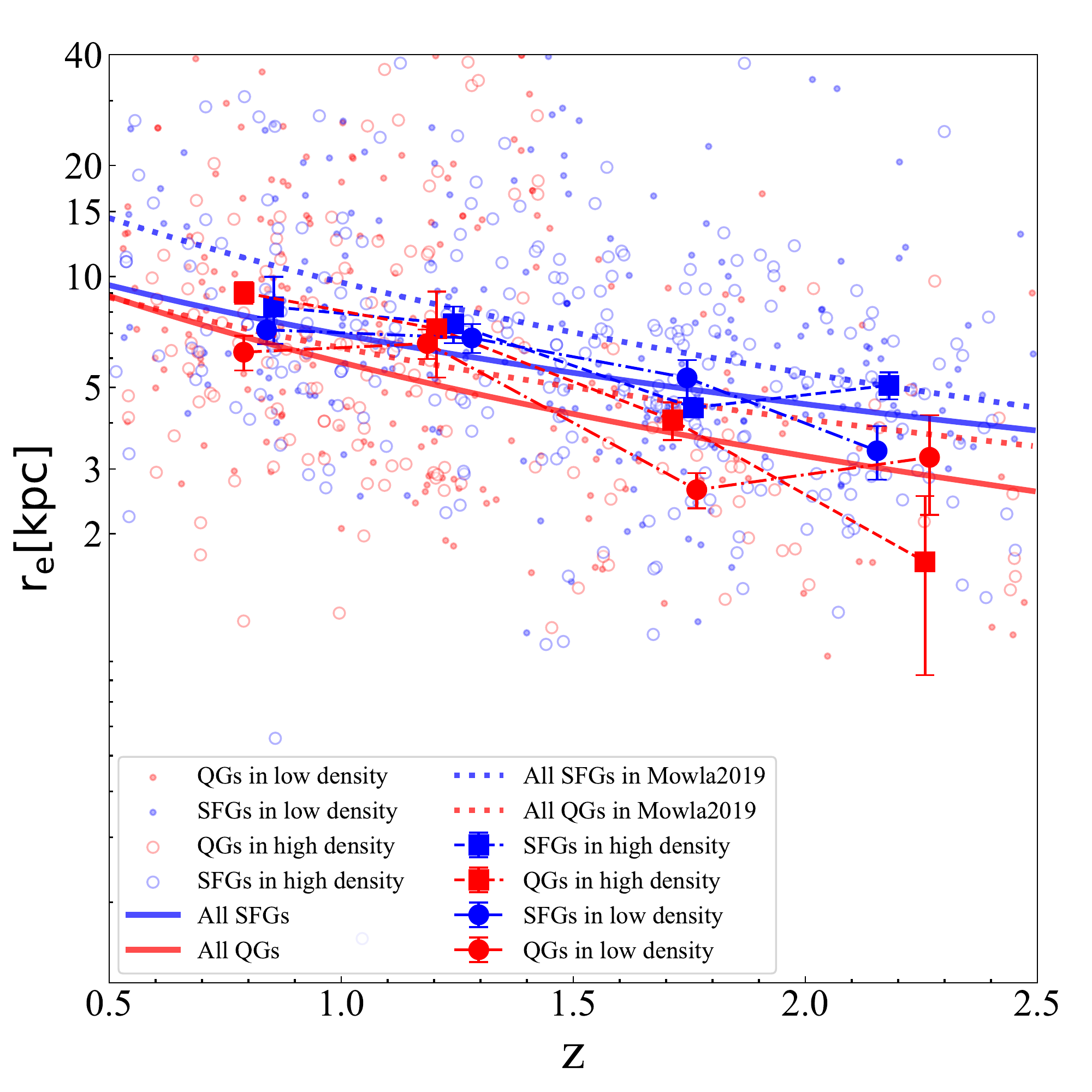}
\caption{Galaxy sizes of the most massive galaxies as a function of redshift for SFGs and QGs. The blue (red) dot and open circles represent individual SFGs (QGs) in the lowest and highest overdensity quarters, respectively. The blue (red) solid line represents the best-fit size evolution of all SFGs (QGs) in our sample. For SFGs (QGs) in the lowest and highest overdensity quarters, we denote their median sizes in each redshift bin as blue (red) solid circles and squares with error bars, respectively.
The error bars are estimated by the bootstrapping method with 1000 times resamplings.
The results from \cite{mowlaCOSMOSDASHEvolutionGalaxy2019} are also shown with blue and red dashed lines for SFGs and QGs, respectively.}
\label{fig5}
\end{figure}

The same as \cite{mowlaCOSMOSDASHEvolutionGalaxy2019}, we also follow \cite{vanderwel3DHSTCANDELSEVOLUTION2014} and correct the size measurements to a common rest-frame wavelength of 5000 \AA. The form of the correction is
\begin{equation}
r_{\rm eff}=r_{\rm eff,F}(\frac{1+z}{1+z_p})^{\frac{\Delta \log\ r_{\rm eff}}{\Delta \log \lambda}}
\end{equation}
where F denotes either $I_{814}$ (for galaxies at $z<1.5$) or $H_{160}$ (for galaxies at $z>1.5$), and $z_p$ is the “pivot redshift” for these respective filters ($z_p$=0.6 for $I_{814}$ and $z_p$=2.2 for $H_{160}$).
Again following \cite{vanderwel3DHSTCANDELSEVOLUTION2014}, the color gradient adopted for star-forming galaxies is
\begin{equation}
\frac{\Delta \log\ r_{\rm eff}}{\Delta \log \lambda}=-0.35+0.12z-0.25log(\frac{M_{\star}}{10^{10}M_{\odot}})
\end{equation}
and for quiescent galaxies it is simply ${\Delta \log\ r_{\rm eff}}/{\Delta \log \lambda}=-0.25$.

The size of the most massive galaxies as a function of redshift is presented in Figure \ref{fig5}. The blue dots and open circles represent SFGs, whereas the red ones represent QGs. As shown in many studies, there is a slight difference between the size--$z$ distributions of these two samples, i.e., SFGs prefer to have a larger size compared to QGs. When considering the size evolution, galaxies are found to be larger at lower redshift. We parameterize the evolution of the median size of star-forming and quiescent galaxies with an empirical function, which is described as:
\begin{equation}
r_{\rm{eff}} = B_z\times(1+z)^{-\beta z}.
\end{equation} 
The best-fit result for SFGs (QGs) with an index of $\beta_z=1.07$ (1.44) is shown as the blue (red) line in Figure \ref{fig5}. 
SFGs seem to have a larger size and a slower size growth compared to QGs, indicating that these two populations may have different mass assemble histories. Stellar mass assembly via star formation might process at all radii in SFGs, but the mode of growth may change after quenching. Dry merger at outer radii may be responsible for the size growth of QGs, which may lead to a faster size evolution.(e.g., \citealt{newmanCANMINORMERGING2012,vandokkumFORMINGCOMPACTMASSIVE2015a}).
 
The results from \cite{mowlaCOSMOSDASHEvolutionGalaxy2019} for galaxies with $M_{\star}>2 \times 10^{11}M_{\odot}$ are also presented in Figure \ref{fig5} for comparison.
In general, the galaxy sizes measured from \cite{mowlaCOSMOSDASHEvolutionGalaxy2019} are remarkably larger than those in this work given a fixing redshift. This difference is reasonable because their sample has a larger stellar mass limit. When we only consider the galaxies with $M_{\star} > 2\times10^{11}M_{\odot}$, there is no significant difference in the median sizes between their work and ours. However, given $\beta_z = 1.40$ and 1.09 for SFGs and QGs in their work, respectively, SFGs seems to have a slight fast evolution at higher stellar mass. But this can also be explained by errors in size measurements in different works. Moreover, \cite{vanderwel3DHSTCANDELSEVOLUTION2014} also reported an evolution with $\beta_z = 0.72$ and 1.24 for SFGs and QGs with $\log(M_{\star}/M_{\odot}) \sim 10.75$, respectively, which is similar like this work that QGs show a  faster size evolution.


However, the main aim of this paper is not to study the size evolution of the most massive galaxies. To investigate the environmental effect on galaxy size, we separate the SFGs into two subsamples according to their local density. In Figure \ref{fig5}, the dots and open circles represent galaxies in the lowest and highest local density quarters, respectively. The blue (red) solid circles and squares with error bars represent the evolution of the median size of SFGs (QGs) in the lowest and highest local density quarters, respectively. The error bars denote the uncertainties of the median values that are estimated by the bootstrapping method with 1000 times resamplings.

As shown in Figure \ref{fig5}, there is no substantial difference in galaxy size distribution between the highest and lowest density quarters for both SFGs and QGs at $1.0<z<2.5$. 
We tested the significance of the size difference by comparing the binned medians between the highest and lowest local density quarters. Such differences in the median sizes are smaller than 3$\sigma$ in all redshift bins for both SFGs and QGs and thus could be neglectable. 
Therefore, we conclude that the environment might have no effect on the galaxy size.

Since galaxy morphology is thought to be determined by both the environment and stellar mass, this may indicate physical mechanisms that determine galaxy shape are more related to stellar mass for these massive galaxies.
Similar results have been reported in many previous studies (e.g., \citealt{retturaFORMATIONEPOCHSSTAR2010,huertas-companyNOEVIDENCEDEPENDENCE2013,kelkarGalaxySizesFunction2015,guEffectEnvironmentStar2021,cutlerDiagnosingDASHCatalog2022}). In the local universe, \cite{huertas-companyNOEVIDENCEDEPENDENCE2013} used a sample from SDSS and did not find any significant size difference for central and satellite early-type galaxies with $11.5<\log(M_{\star}/M_{\odot})<11.8$. At higher redshift, \cite{retturaFORMATIONEPOCHSSTAR2010} found that early-type galaxies in a cluster at $z\sim 1.23$ follow comparable mass versus size relation with those in the field. Based on a sample at $z<1.2$ in the COSMOS-DASH field, \cite{cutlerDiagnosingDASHCatalog2022} obtained the galaxy density from the COSMOS density field catalog (\citealt{darvishCOMPARATIVESTUDYDENSITY2015,darvishCosmicWebGalaxies2017b}) and reported that galaxy size is constant with density for all mass bins. In this work, we extend their result to higher redshift.

\begin{figure*}[htb!]
\centering
\includegraphics[width=0.9\textwidth, height=0.58\textwidth]{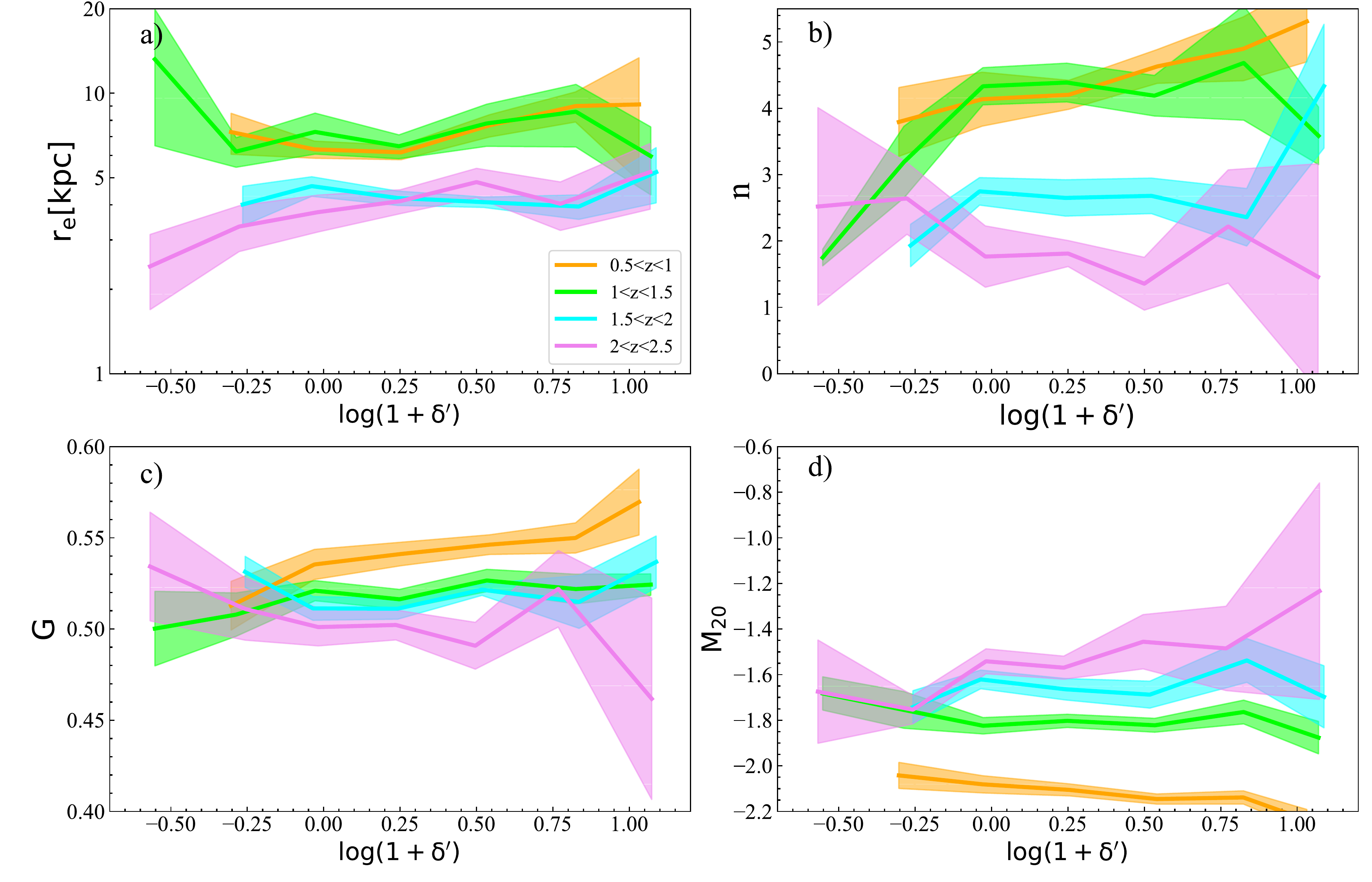}
\caption{Galaxy structure parameters, $r_{\rm e}$ (Panel a), $n$ (Panel b), $G$ (Panel c), and $M_{20}$ (Panel d), as a function of the overdensity at redshift $0.5<z<2.5$. In each panel, the orange, lime, aqua, and violet lines represent the median values in different redshift bins, while the corresponding shaded areas indicate the $1\sigma$ uncertainties of the medians over each overdensity bin estimated via the bootstrapping method.}
\label{fig6}
\end{figure*}

\begin{table*}[htbp]
	\centering
	\caption{Spearman correlation coefficients between the morphological parameters and overdensity in different redshift bins.}
	\label{tab1}  
	\begin{tabular}{ccccc}
		\hline\hline\noalign{\smallskip}	
		Redshift range & $r_{\rm e}$ & $n$ & $G$ & $M_{20}$  \\
		\noalign{\smallskip}\hline\noalign{\smallskip}
		  $0.5<z<1.0$ & $0.16\pm 0.05$ & $0.14\pm 0.05$ & $0.09 \pm 0.05$ & $-0.04 \pm 0.05$ \\
		$1.0<z<1.5$ & $0.02\pm 0.07$ & $0.06 \pm 0.04$ & $0.09\pm 0.04$ & $0.02 \pm 0.04$  \\
		$1.5<z<2.0$ & $0.02\pm 0.05$  & $0.07 \pm 0.05$ & $0.04\pm 0.05$ & $-0.05 \pm 0.05$ \\
		$2.0<z<2.5$ & $0.20\pm 0.07$ & $-0.01\pm 0.08$ & $-0.06\pm 0.08$ & $0.10 \pm 0.08$ \\
		\noalign{\smallskip}\hline
	\end{tabular}
\end{table*}

Furthermore, our sample's structure parameters are shown as functions of overdensity in redshift bins with $\Delta z=0.5$ in Figure \ref{fig6}. Within each redshift bin, we divide our sample into bins of overdensity (with an interval $\Delta \log(1+\delta') =0.3)$) and compute the medians of the binned distributions (solid lines) and the corresponding $1\sigma$ uncertainties (shaded areas) via the bootstrapping method. From this figure, it is clear that galaxy morphologies are redshift-dependent. From Panels a) and b), galaxies at lower redshift tend to be more bulge-dominated with a higher S\'{e}rsic index and have a larger effective radius. But when considering the relation between morphology and overdensity, it seems that there is no significant environmental dependence for morphology. As for the nonparametric statistics, Panels c) and d) in Figure \ref{fig6} reflect that, similar to $n$ and $r_{\rm e}$, 
there are also no significant changes in the medians of both the Gini index and $M_{20}$ when $\log(1+\delta')$ varies from -0.5 to 1. To get a statistical result, the Spearman rank correlation coefficients between the structure parameters and the local density are calculated and listed in Table \ref{tab1}. The 1$\sigma$ uncertainties of the correlation coefficients are also derived via the bootstrapping method. Clearly, the Spearman correlation coefficients between morphology and environment are all less than 0.3, suggesting that the environmental effects on these massive galaxies may not change their shape.

However, from the figure, there is weak evidence that galaxies at $0.5<z<1.0$ in the highest density environment seem to have a larger S\'{e}rsic index, which is supported by the corresponding non-zero (within 3$\sigma$) Spearman correlation coefficient given in Table \ref{tab1}. A higher rate of (minor) mergers may be invoked to explain the environmental dependence on galaxy morphologies. 
Similar observations have also been reported in \cite{kawinwanichakijEffectLocalEnvironment2017} and \cite{guEffectEnvironmentStar2021}, in which the authors found weak evidence that the S\'{e}rsic index of QGs in the highest density environment is larger than that of galaxies in the lowest environment at $0.5<z<1.0$. However, when they consider the statistical significance, the difference is within 2$\sigma$. The catalogs they used mainly come from the CANDELS fields, which are pretty small samples given the limited sky area. We attempt to get a more reliable statistical result with a larger sample from COSMOS-DASH, but the result is still weak, suggesting that the current sample we use might still not be large enough. Further deep field surveys, such as the one scheduled by the Chinese Space Station Telescope, are expected to enlarge the sample size of the most massive galaxies and thus reach a more reliable conclusion.

\section{Environmental effect on star forming activity} \label{sec:strformation effect}

Besides galaxy morphologies, galaxy star formation state is also an important indicator in studying galaxy evolution history. It has been shown in many previous studies that the environment has an important effect on galaxy properties. For less massive galaxies, \cite{baloghGalaxyEcologyGroups2004}, \cite{kauffmannEnvironmentalDependenceRelations2004a}, and \cite{baldryGalaxyBimodalityStellar2006} have shown that the galaxies at denser environments have lower median sSFR at lower redshift ($z<0.2$). However, here is no consensus for galaxies at intermediate redshift ($z \sim 1$). Some studies showed that the environment has the same effect as in local universe \citep{muzzinGEMINICLUSTERASTROPHYSICS2012a}, while some studies argued that sSFR is independent of environment \citep{scovilleEVOLUTIONGALAXIESTHEIR2013}. As for higher redshift, \cite{darvishEFFECTSLOCALENVIRONMENT2016} found that there is no environmental dependence of sSFR at redshift $1.1<z<3.1$.

\begin{figure}[htb!]
\centering
\includegraphics[width=0.48\textwidth, height=0.48\textwidth]{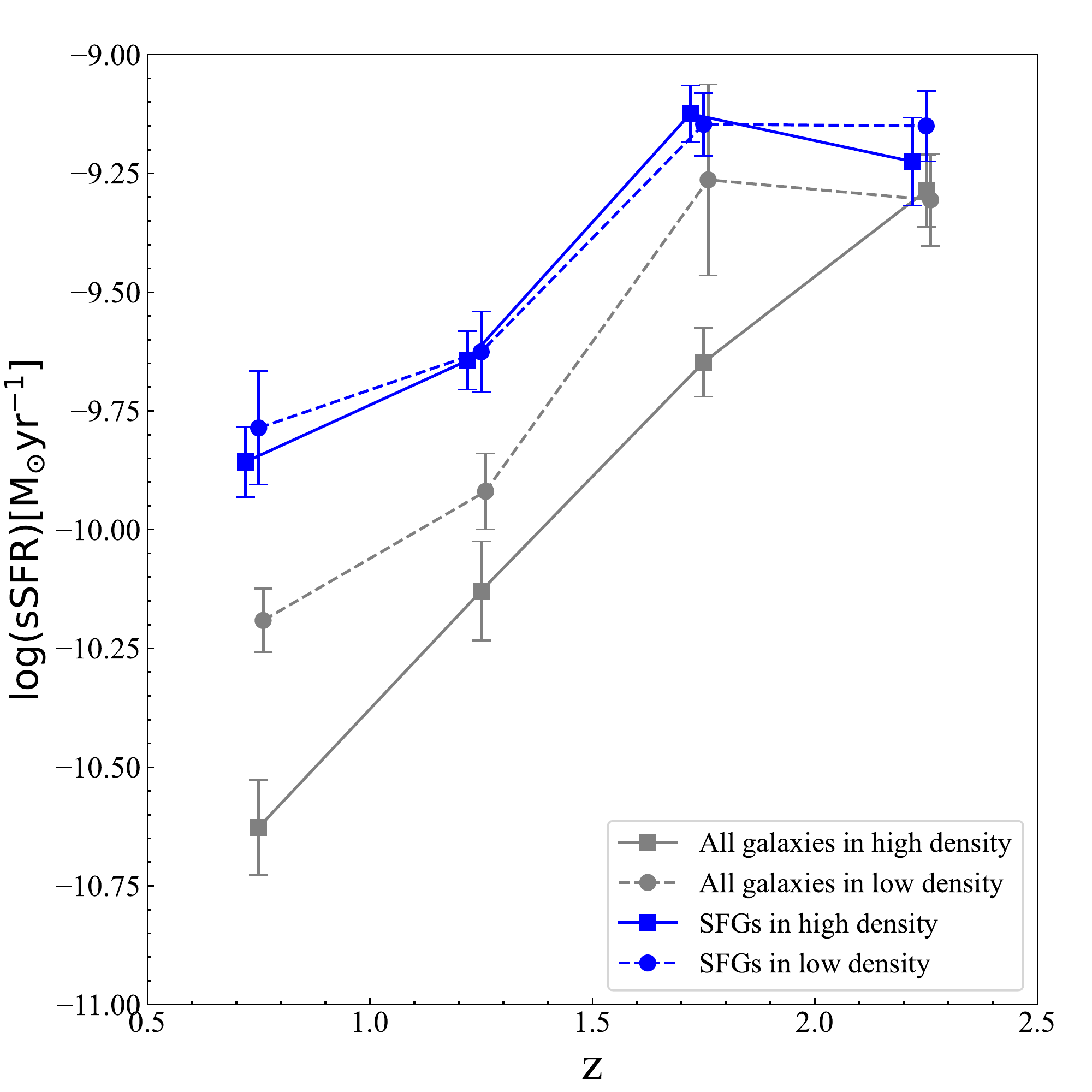}
\caption{Median sSFR as a function of redshift in different environments for all galaxies (grey) and SFGs only (blue).
The circles connected by dashed lines and squares connected by solid lines represent galaxies in the lowest and highest local density quarters, respectively. The error bars indicate the $1\sigma$ uncertainties of the medians estimated via the bootstrapping method. The sSFR shows an environmental dependence for all galaxies but becomes independent of the environment at all redshifts when we consider the SFGs only.}
\label{fig7}
\end{figure}

In this section, we will study whether the environment has an effect on the sSFR of the most massive galaxy. In Figure \ref{fig7}, we show the median sSFR of galaxies in the lowest and highest density quarters as a function of redshift.
As shown in many previous studies, galaxies tend to have an active star formation state at higher redshift regardless of the environment, which may lead to a larger sSFR \citep{darvishEFFECTSLOCALENVIRONMENT2016}. We note that sSFR at $2.0<z<2.5$ is similar to that at $1.5<z<2.0$ in low density, which might be a bias due to the relatively small number of galaxies in this redshift range. A significantly lower sSFR can be found in denser environments at $0.5<z<2.0$, suggesting that different from galaxy morphologies, the sSFR of these massive galaxies is affected by the local density at least at $0.5<z<2.0$. But within the redshift range of $2.0<z<2.5$, the sSFR seems to have no dependence on overdensity.

Comparing the environmental effect on sSFR between the most massive galaxies and less massive galaxies, we can find the environmental dependence of sSFR for the most massive galaxies occurs earlier ($z\sim 2.0)$ than less massive galaxies ($z\sim 1$; \citealt{muzzinGEMINICLUSTERASTROPHYSICS2012a}). This might be partly due to the different environmental quenching mechanisms for the most massive and less massive galaxies. However, the role of the environment in the evolution of the most massive galaxies is still in debate.
Based on galaxies with $\log(M_{\star}/M_{\odot})>10.7$ from the COSMOS field, \cite{darvishEFFECTSLOCALENVIRONMENT2016} found that the median sSFR of most massive galaxies at denser environment seems to be lower at $z\lesssim 1.1$, but at higher redshift, the environment seems to have no effect on galaxy star formation state. But in \cite{Chartab_2020}, the most massive galaxies ($\log(M_{\star}/M_{\odot})>11$) in the CANDELS fields are found to have lower SFR in higher density environment at $0.1<z<3.1$. The reason for this discrepancy might be the cosmic variance and/or how the environment is traced. Larger and deeper spectroscopic and/or photometric surveys are still needed to resolve this issue. 

The difference in sSFR between different environments may be caused by the reasons that (1) star formation is suppressed in the highest density quarter, and (2) there are more QGs in the highest density quarter. To investigate which one is dominant, we examine whether there is a difference in sSFR between SFGs in the highest and lowest density quarters in Figure \ref{fig7}, in which the binned medians of sSFR for SFGs in different environments are also denoted by blue symbols as a function of redshift.

At $0.5<z<1.0$, the sSFR of SFGs in the lowest local density quarter is slightly larger than in the highest quarter. The difference between these two subsamples is insignificant when either the uncertainties are considered, or a K-S test is performed.
The environment might affect the sSFR for SFGs at this redshift range, but the effect should be very weak. As for higher redshift, the environment may not play a critical role. Similar results have also been shown in \cite{darvishEFFECTSLOCALENVIRONMENT2016}, once the stellar mass and star formation state are fixed, SFR for galaxies at $0.1<z<3.1$ is independent of environment.

\begin{figure}[htb!]
\centering
\includegraphics[width=0.48\textwidth, height=0.48\textwidth]{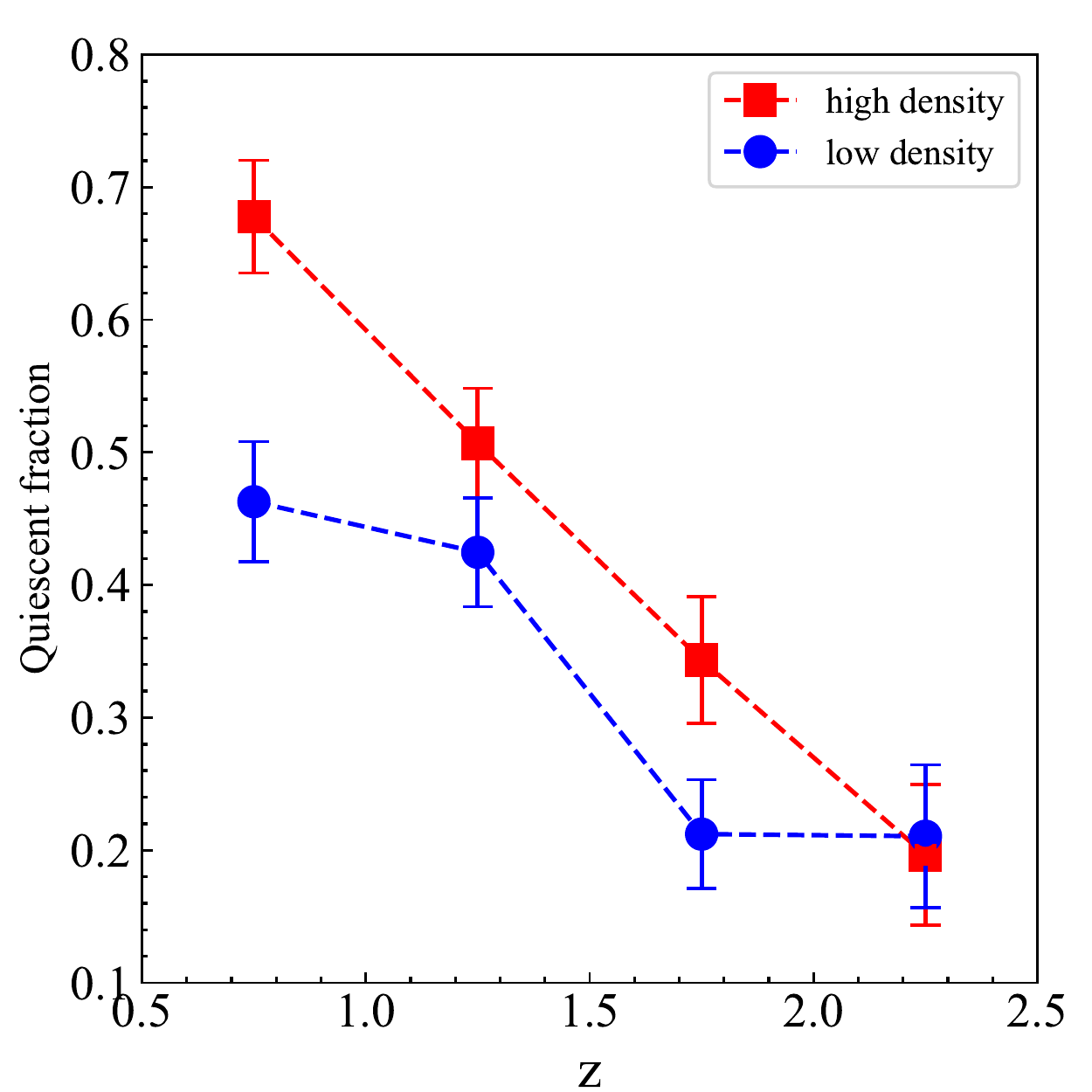}
\caption{Fraction of QGs as a function of redshift in different overdensity bins. The blue circles and red squares represent the median quiescent fractions for galaxies in the lowest and highest density quarters, respectively. The error bars are the 1$\sigma$ uncertainties of the medians derived via the bootstrapping method. We find that the fraction of QGs shows a clear dependence on the overdensity at $z<2$. However, we do not find a variation of the quiescent fraction with the environment at higher redshift ($z>2$).}
\label{fig8}
\end{figure}

To check the second reason, we present the quiescent fraction (i.e., the fraction of QGs) as a function of redshift in different overdensity bins in Figure \ref{fig8}. According to this figure, at $0.5<z<2.0$, there is a significant gap in the quiescent fractions between the galaxies in different overdensity quarters. This indeed supports the idea that the environment has played an effective role in galaxy quenching at $0.5<z<2.0$. But at $z>2.0$, the gap between these two overdensity quarters disappears. This is consistent with the result from Figure \ref{fig7} that the environmental effect on the most massive galaxies could be neglected at higher redshift.

With the data from the COSMOS survey, \cite{darvishEFFECTSLOCALENVIRONMENT2016} found that galaxies residing in higher local density seem to have a higher quiescent fraction with stellar mass ($\log(M_{\star}/M_{\odot})>10.8$). But in the study of \cite{paulino-afonsoVISCOSSurvey2018}, with a cluster at $z\sim 0.84$, they found a nearly constant quiescent fraction of $\sim30\%$ for galaxies with $\log(M_{\star}/M_{\odot})>10.75$, indicating no dependence on the local density.
This difference may be due to the smaller sample size of \cite{paulino-afonsoVISCOSSurvey2018}.

Our results have revealed some indications related to the quenching mechanism of the most massive galaxies.
In Figure \ref{fig7}, we find that the sSFR for SFGs shows little dependence on the local environment. In other words, the local environment has no significant effect on the star formation activity for SFGs. This agrees well with a plethora of observations and simulations at different redshifts that many properties of SFGs that are directly or indirectly linked to star formation activity (e.g., SFR, sSFR) do not depend on their host environment (e.g., \citealt{patelDEPENDENCESTARFORMATION, hayashiMappingLargescaleStructure2014,duivenvoordenHELPStarFormation2016}). Therefore, the reason for a lower sSFR for all galaxies in the denser environment should be the higher quiescent fraction compared to those in the lower-density environment, which is demonstrated in Figure \ref{fig8}. \cite{faisstConstraintsQuenchingMassive2017} showed that major merger is the dominant evolutionary agent to produce massive, passive galaxies by analyzing the size--mass relation of QGs. Since the environment has no significant effect on the sSFR of SFGs, but can lead to a higher quiescent fraction at $z<2$, we suspect that the environmental quenching process should happen in a short timescale. One possible physical mechanism that is able to interpret our observations is the merger. In denser environments, galaxies are more likely to merge with other galaxies, leading to a higher quiescent fraction.

\section{Summary} \label{sec:summary}

Based on the overdensity maps derived by the Bayesian metric for the COSMOS-DASH fields, we obtain a large sample of the most massive galaxies covering a wide redshift range to investigate the environmental effect on galaxy star formation state and morphology. The main conclusions are as follows.

(1) Different from the less massive galaxies, the morphologies of these massive galaxies are independent of the local environmental overdensity. But at $0.5<z<1.0$, there is weak evidence that S\'{e}rsic of galaxies in higher density seems to be larger, which might be caused by mergers.

(2) For our massive galaxies, the median sSFR of all galaxies is affected  by the environment at $0.5<z<2.0$ and becomes almost independent of the environment at $2.0<z<2.5$. When we only consider SFGs, their median sSFR does not significantly change with the environment at all redshift, indicating the environmental independence of the main sequence of SFGs.

(3) However, the chance of a galaxy becoming quiescent at $z<2.0$ increases in denser environments. Hence, the high-density environment may promote the transformation of star-forming galaxies into quiescent galaxies. Combined with the environmental independence of the median sSFR of SFGs, the environmental quenching process may be a merger.

The relationship between galaxy properties and the local environment still needs further study, especially at higher redshift. Thanks to the successful launch of JWST, we will be able to obtain deeper galaxy images with high resolution in the NIR bands. Furthermore, the upcoming Chinese Space Station Telescope plans to carry out an optical deep field survey covering an unprecedentedly large sky area, which is expected to provide a pretty large sample of the most massive galaxies in the near future. These surveys will help us to better understand the morphology transformation and, thus the early evolution of the most massive galaxies at higher redshift.

\begin{acknowledgments}
This paper is based on observations made with the NASA/ESA HST, obtained at the Space Telescope Science Institute, which is operated by the Association of Universities for Research in Astronomy, Inc., under NASA contract NAS 5-26555. These observations are associated with program HSTGO-14114. Support for GO-14114 is gratefully acknowledged.
This work is supported by the Strategic Priority Research Program of Chinese Academy of Sciences (Grant No. XDB 41000000), the National Science Foundation of China (NSFC, Grant No. 12233008, 11973038), the China Manned Space Project (No. CMS-CSST-2021-A07) and the Cyrus Chun Ying Tang Foundations.
Z.S.L. acknowledges the support from the China Postdoctoral Science Foundation (2021M700137). Y.Z.G. acknowledges support from the China Postdoctoral Science Foundation funded project (2020M681281).
\end{acknowledgments}

\bibliography{ref}
\bibliographystyle{aasjournal}

\end{document}